\documentclass[a4paper, amsfonts, amssymb, amsmath, reprint, showkeys, nofootinbib, twoside, superscriptaddress]{revtex4-1}
\usepackage[english]{babel}
\usepackage[utf8]{inputenc}
\usepackage[colorinlistoftodos, color=green!40, prependcaption]{todonotes}
\usepackage{amsthm}
\usepackage{mathtools}
\usepackage{physics}
\usepackage{xcolor}
\usepackage{graphicx}
\usepackage[left=23mm,right=13mm,top=35mm,columnsep=15pt]{geometry} 
\usepackage{adjustbox}
\usepackage{placeins}
\usepackage[T1]{fontenc}
\usepackage{lipsum}
\usepackage{csquotes}
\usepackage[pdftex, pdftitle={Article}, pdfauthor={Author}]{hyperref} 
\begin{document}
\title{Electron Holographic Mapping of Structural and Electronic
Reconstruction at Mono- and Bilayer Steps of h-BN}

\author{Subakti~Subakti}
    \affiliation{Leibniz Institute for Solid State and Materials Research Dresden, Helmholtzstraße 20, 01069 Dresden, Germany}

\author{Mohammadreza~Daqiqshirazi}
	\affiliation{Chair of Theoretical Chemistry, Technische Universität Dresden, Bergstraße 66, 01069 Dresden, Germany}

\author{Daniel~Wolf}
	\affiliation{Leibniz Institute for Solid State and Materials Research Dresden, Helmholtzstraße 20, 01069 Dresden, Germany}
	
\author{Martin~Linck}
    \affiliation{Corrected Electron Optical Systems GmbH, Englerstrasse 28, D-69126 Heidelberg, Germany}
    
 \author{Felix~L.~Kern}
	\affiliation{Leibniz Institute for Solid State and Materials Research Dresden, Helmholtzstraße 20, 01069 Dresden, Germany} 
\author{Mitisha Jain}

\affiliation {Institute of Ion Beam Physics and Materials Research, Helmholtz-Zentrum Dresden-Rossendorf, Bautzner Landstraße 400, 01328 Dresden, Germany}
\author{Silvan Kretschmer}

\affiliation {Institute of Ion Beam Physics and Materials Research, Helmholtz-Zentrum Dresden-Rossendorf, Bautzner Landstraße 400, 01328 Dresden, Germany}	
	
\author{Arkady V. Krasheninnikov}
\affiliation {Institute of Ion Beam Physics and Materials Research, Helmholtz-Zentrum Dresden-Rossendorf, Bautzner Landstraße 400, 01328 Dresden, Germany}
\affiliation
{Department of Applied Physics, Aalto University, P.O. Box 11100,
FI-00076 Aalto, Finland}
	
\author{Thomas~Brumme}
    \email{thomas.brumme@tu-dresden.de}
	\affiliation{Chair of Theoretical Chemistry, Technische Universität Dresden, Bergstraße 66, 01069 Dresden, Germany}

\author{Axel~Lubk}
    \email{a.lubk@ifw-dresden.de}
	\affiliation{Leibniz Institute for Solid State and Materials Research Dresden, Helmholtzstraße 20, 01069 Dresden, Germany}
	\affiliation{Institute of Solid State and Materials Physics, Technische Universität Dresden, Dresden, Germany}

\date{\today} 

\begin{abstract}

Here, by making use of medium and high resolution autocorrected off-axis electron holography, we directly probe the electrostatic potential as well as in-plane and out-of-plane charge delocalization at edges and steps in multilayer hexagonal boron nitride. In combination with ab-initio calculations, the data allows to directly reveal the formation of out-of-plane covalent bonds at folded zig-zag edges and steps comprising two monolayers and the absence of which at monolayer steps. The technique paves the way for studying other charge (de)localization phenomena in 2D materials, e.g., at polar edges, topological edge states and defects.
\end{abstract}


\maketitle

\section{Introduction} \label{sec:Introduction}

Two-dimensional (2D) materials consisting of a single or a few atomic layer(s) have been intensively studied for their intriguing chemical and physical properties \cite{Das2021,Li2018}. Particular focus has been put on 2D van der Waals (vdW) materials with honeycomb lattice such as graphene \cite{Novoselov2004}, hexagonal boron nitride (h-BN) \cite{Zhang2017}, and transition metal dichalcogenides (TMDCs)  \cite{Manzeli2017}. Here, exfoliation allows fabrication of various vdW homo/hetero-structures\cite{Geim2013}, which facilitate band-structure engineering \cite{Withers2015,Nakamura2020}, proximity-induced exchange coupling \cite{Zollner2020}, or gate tunable photovoltaic devices\cite{Flory2015}, amongst others.

Despite the fact that graphene has sparked the rapidly growing 2D materials (2DMs) research field, h-BN has emerged as a key material for 2D vdW heterostructure building blocks \cite{Zhang2017, Nakamura2020, Zheng2021, Roy2021}. This mainly derives from its stable honeycomb lattice (in-plane lattice constant 2.504 \AA), stabilized by strong in-plane sp2-hybridized covalent bonds between alternating boron (B) and nitrogen (N). By virtue of its relatively weak interplanar vdW  forces (3.33 \AA\,  interplanar distance) it is possible to obtain dangling-bond free exfoliated surfaces as a substrate, as passivation layers, or as insulating layers in the vdW heterostructures \cite{Roy2021, Wang2017}. 

Amongst others, future developments within the field crucially depend on understanding and tailoring the one-dimensional edges and surface steps of 2DMs, because of their unique properties and their prominent role in the 2DMs' chemical and physical properties, such as their mechanical and chemical stability. In the particular case of monolayer honeycomb 2D materials two distinct variants of edges, referred to as zigzag and armchair, exist, which can exhibit very different electronic properties \cite{Alem2009,Yamanaka2017}. Ab-initio calculations predict, for instance, that armchair edges of h-BN monolayers are insulating and non-magnetic, whereas zigzag edges are metallic and ferromagnetic. Furthermore, exfoliated 2DMs typically exhibit ripples and wrinkles, crystallographic point and line defects \cite{Park2020}, a varying number of stacked layers\cite{Gilbert2019}, and stacking layer mismatches, which can also modify the electronic properties. For example, atomically sharp twin boundaries in h-BN show conducting behaviour with a zero bandgap \cite{Park2020}. Similarly, crystallographic point defects can serve as color-centers and hence single photon emitters \cite{Yim2020, Fischer2021,Koehl2011} or can be manipulated to control functional properties such as thermoelectricity \cite{Wu2020}. 

\begin{figure*}
		\centering
		\includegraphics[width=0.95\linewidth]{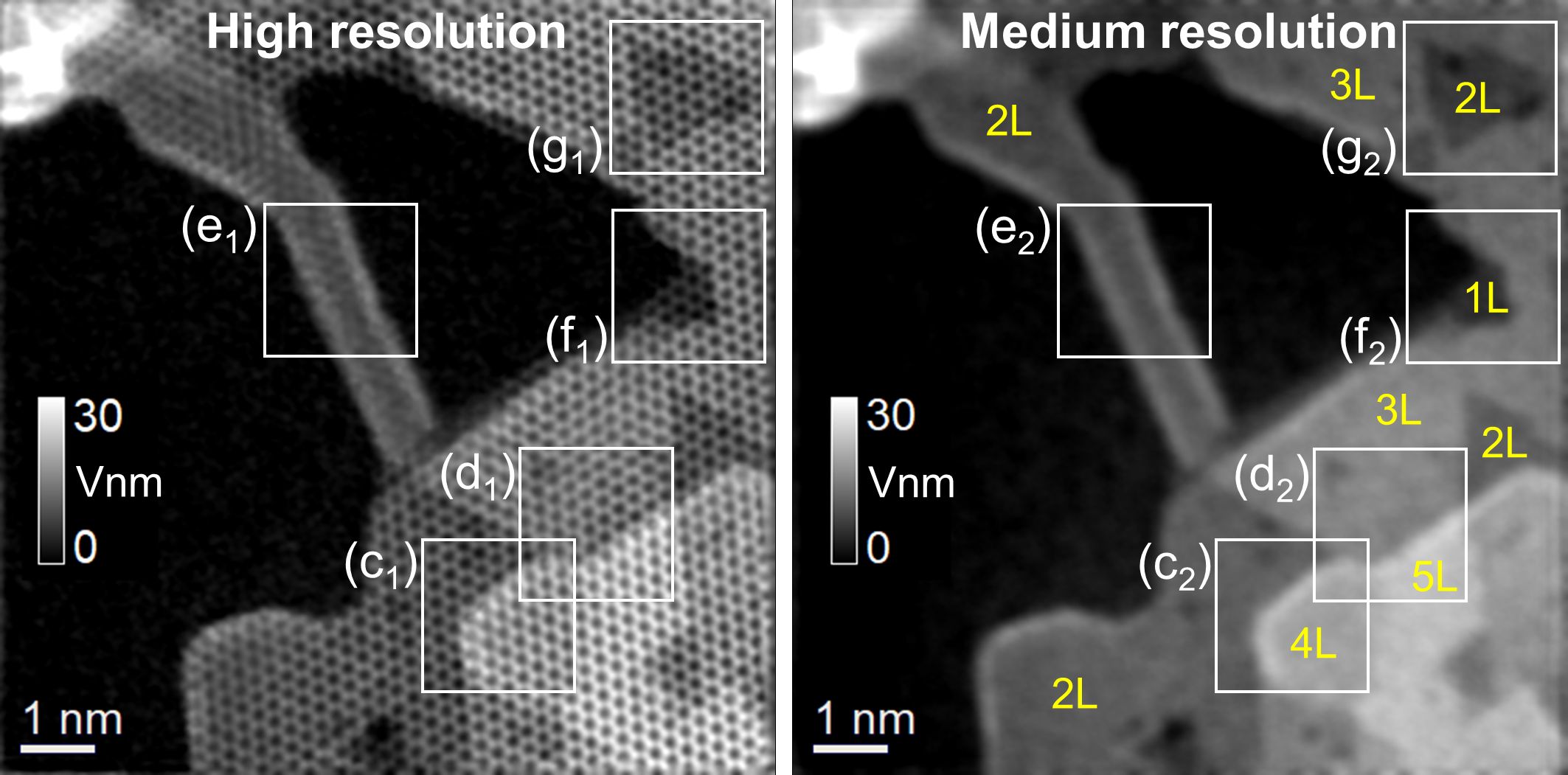}
		\caption{Atomic-scale (high resolution) and nanometer-scale (medium resolution)  projected potential of multilayer h-BN reconstructed from off-axis electron holography experimental data. The local number of layers is indicated in yellow (e.g., 3 L). The white squares show the regions of interest to be analyzed in detail in Fig.~\ref{Fig:Analysis}.
		}
		\label{Fig:Holo_Data}
\end{figure*}

Probing and studying these properties requires atomic resolution microscopy techniques, such as scanning probe microscopy techniques, which allow resolving the atomic structure and are sensitive to electronic properties at the same time. Transmission electron microscopy (TEM) based techniques have been successfully used to study freestanding 2DMs including, e.g., defects and edges of freestanding 2DMs \cite{Alem2009}. Going one step further, i.e., revealing electronic properties, of freestanding edge and step structures with TEM has been proven difficult (see \cite{Alem2012} for a notable early exception), mostly because of the small elastic and inelastic scattering cross-section of 2DMs. Recently off-axis holographic and 4D STEM techniques have been successfully used to study various 2DMs exploiting the proportionality between projected electrostatic potential of 2DMs and the phase modulation of the scattered electron wave \cite{Ortolani2011, Cooper2014, Winkler2018, Fang2019}. 

In Ref. \cite{Kern2020}, we established how the small elastic scattering cross-sections and the ensuing Fourier space symmetries of the scattered electron wave can be exploited to automatically correct for residual aberrations and hence retrieve accurate electrostatic potentials of 2DMs at atomic resolution. Since the electrostatic potential is closely related to the charge density of investigated materials, it is also sensitive to charge redistributions at point defects and nanoscale deformations including corrugated surfaces and wrinkling. 
In the following, we apply autocorrected off-axis holography to reveal structural and electronic reconstruction at edges and steps of exfoliated multilayers of h-BN. We make use of the link between charge density and potential, which is correlated to electronic properties to be computed by ab-initio electronic structure calculations. Using this machinery, we can directly reveal the formation of out-of-plane covalent bonds at the zig-zag edges of h-BN comprising bilayers and the absence of such a significant structural and electronic reconstructions in monolayer edges and steps.

\section{Experimental and Theoretical Methods} \label{sec:Experiment}

\begin{figure*}
		\centering
		\includegraphics[width=1\linewidth]{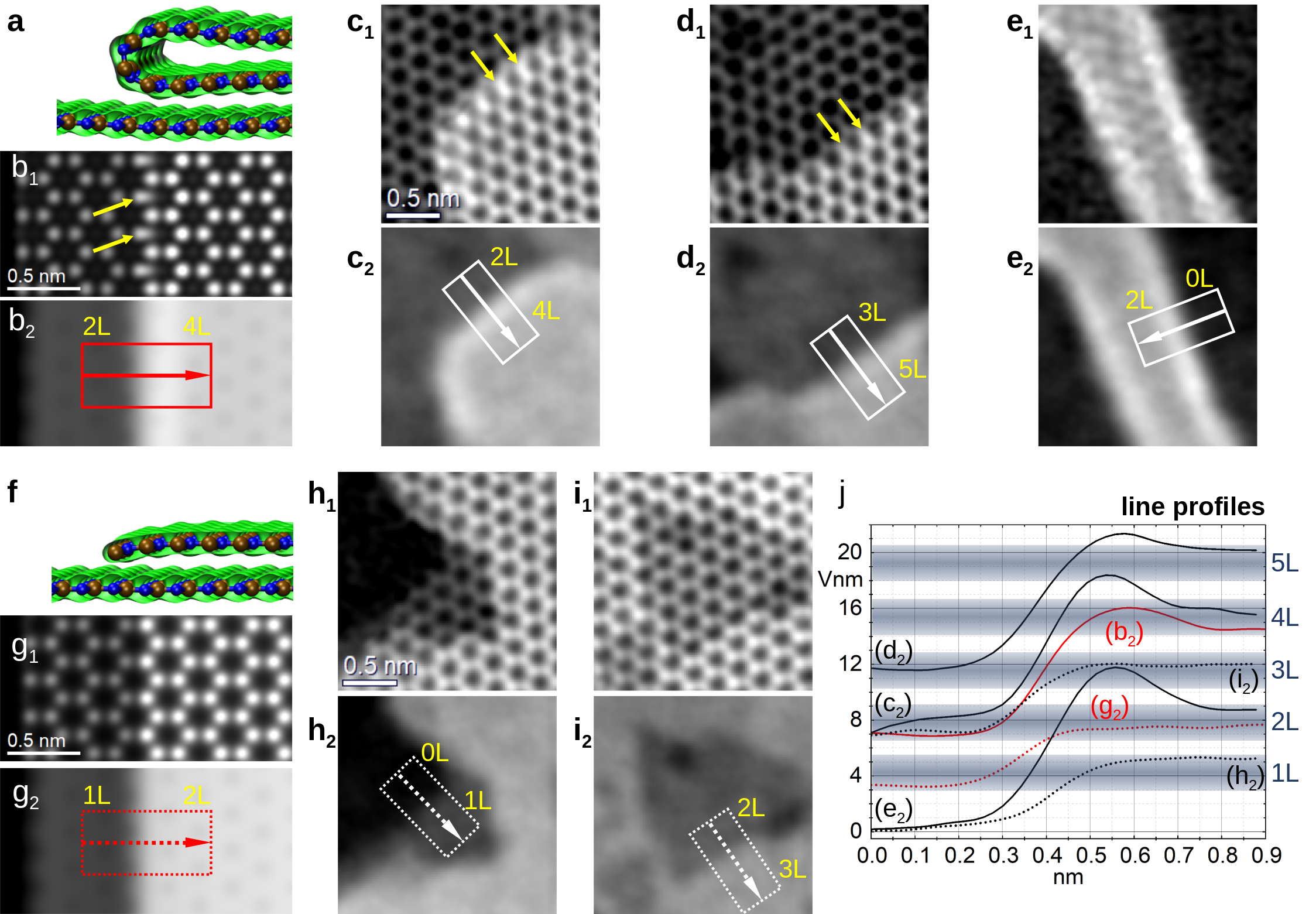}
		\caption{Projected potential analysis at h-BN mono- and bilayer steps. (a) Simulated relaxed structure (blue: N, brown: B) and electron density (green isodensity surface) from DFT calculations for bilayer steps. (b) projected potentials of the simulated structure, where the subscript numbers 1 and 2 distinguish between atomic and medium resolution, respectively. (c-e) Zoomed-in views of the experimental data indicated by white squares in Fig.~\ref{Fig:Holo_Data}. The  number of monolayers L as determined from projected potentials are indicated. (f-i) Same as (a-e), but for monolayer steps. (j) Line profiles along line scans indicated by arrows in (b-e) and (g-i), where the rectangles illustrate the integration width perpendicular to the scan direction. The theoretical and experimental data are plotted in red and black, respectively. The line profiles of the two-layer/monolayer steps are plotted as solid/dashed line. The number of monolayer is visualized as grey-blue bands in the diagram to consider for experimental uncertainties (e.g., slight charging, instabilities).}
		\label{Fig:Analysis}
\end{figure*}

As described in Ref.~\cite{Kern2020}, auto-corrected off-axis electron holography allows to quantitatively retrieve projected electrostatic potentials of 2DMs at atomic resolution. Thin h-BN flakes comprising several multilayers have been prepared on a Quantifoil TEM grid using the multistep exfoliation process described in Ref. \cite{Alem2009}. Subsequently, the multilayer flake under consideration was further thinned inside the TEM by an electron-beam shower of about 20 nA. This procedure created a terrace landscape of increasingly thinner h-BN layers eventually reaching the monolayer limit (Fig. \ref{Fig:Holo_Data}). Utilizing this in-situ thinning in the high vacuum of a TEM column prevented the saturation or decoration of as-prepared steps and edges with foreign atoms, notably hydrogen. Moreover, perfect AB stacking order (i.e., N on top of B) of bulk h-BN is preserved within the terraces.

The following in-depth analysis of edges and steps in h-BN is based on the projected potential dataset reconstructed in Ref. \cite{Kern2020} (Fig. \ref{Fig:Holo_Data}). It readily reveals h-BN terraces with thicknesses ranging from 1-6 monolayers and corresponding edges and steps, which were deliberately generated by the electron beam irradiation in the TEM as described previously. We refer the reader to Ref.~\cite{Kern2020} for the details of the holographic experiment, the reconstruction, and the autocorrection of aberrations including the defocus.

We analyze the projected potential data at two spatial resolutions, the as-reconstructed atomic resolution (Fig. \ref{Fig:Holo_Data}b) and at medium resolution (Fig. \ref{Fig:Holo_Data}a), where only the average potential per unit cell is considered. The latter was generated from the high-resolution dataset by convolution with a kernel of the size of a unit cell. As detailed in Ref.~\cite{Kern2020}, the projected potential averaged over some prismatic volume of base area $S$ and height $t$ (corresponding to sample thickness) defined though $\bar{\Phi}=\frac{1}{S}\int_{S\times t} \Phi \left( \boldsymbol{r} \right) d^3r $ approximately scales with the sum over the second spatial moments $\langle r_n^2 \rangle = \int \rho_n r_n^2 dr_n$ of the atomic charge densities $\rho_n$ (atomic index $n$) within the averaged region, i.e., $\bar{\Phi}\propto\sum_n\langle r_n^2 \rangle$, and hence serves as a fingerprint for the charge delocalization (i.e., larger delocalization leads to higher second moment and hence higher average potential). Moreover, the average projected potential allows to directly count the number of layers (shown in Figs. \ref{Fig:Holo_Data} and \ref{Fig:Analysis}), which we use to classify the edges and steps into monolayer and bilayer edges and steps (others are not present). Note that all of them terminate in the zig-zag structure, which is the energetically favorable configuration in h-BN. Finally, we generated a set of edge and step zoom-ins facilitating a detailed comparison to first principle calculations, described subsequently.

To describe the mono- and bilayer steps in h-BN an all-electron formalism based on density functional theory (DFT) as implemented in FHI-aims \cite{Blum2009} was employed together with the Perdew–Burke–Ernzerhof (PBE) exchange-correlation functional.\cite{Perdew1996} Additionally, the van der Waals correction based on the Tkatchenko-Scheffler method \cite{Tkatchenko2009} was used. All step structures were embedded in supercells comprising 120 atoms, which were relaxed using tight tier 1 numeric atom-centered orbitals until the forces were smaller than $10^{-3}$\:eV/\AA{} and the electron density changed by less than $10^{-6}$\:e/\AA. The different steps were simulated employing a supercell geometry with about 60 \AA{} vacuum perpendicular to the layers and 30 \AA{} between the periodically repeated edges. The starting structures prior to relaxation have been generated by placing one flat finite h-BN monolayer terminating in a zig-zag edge over a continuous h-BN layer in case of the monolayer step and a flat finite h-BN double layer terminating in a zig-zag edge in case of the bilayer step. The different layers were always stacked in the bulk stacking order (i.e., alternating B and N atoms along $z$). In case of the monolayer step, we additionally mirrored the structure in out-of-plane direction in order to avoid the built up of a polarization potential in the otherwise polar step structure.  We used a $\Gamma$-centered Monkhorst-Pack \cite{Monkhorst1976} k-point grid of $13\times1\times1$ points and interrupted layers (see results further below) included scalar relativistic corrections (ZORA). For the graphical presentation of the charge density with VMD,\cite{Humphrey1996} voxels with a maximum of 0.2 \AA{} edge length were utilized.

To understand the morphology of the edges of h-BN sheets under the experimental conditions, the displacement rates of atoms under electron beam were assessed using the DFT-based molecular dynamics (MD) and the McKinley-Feshbach formalism \cite{McKinley1948}, as done previously for various 2D materials \cite{Meyer2012}. 
The DFT-MD simulations were carried out using the Vienna ab-initio Simulation Package (VASP) \cite{Kresse1993,Kresse1996,Kresse19962,Kresse1999}. The Perdew-Burke-Ernzerhof (PBE) exchange-correlation functional was used \cite{Perdew1997}. The edge of h-BN sheet was modelled
as a h-BN ribbon with a width of about 16 {\AA}. The dangling bonds at the edges of the ribbon were saturated with hydrogen atoms. The periodic images of the supercell were separated by 9 \AA in the inplane and by 40 \AA in the transverse directions. Tight energy convergence (10$^{-5}$ eV) criteria was used for electronic steps on a  $1\times1\times1$ $\Gamma$-centered kpoint mesh. The cutoff energy was set as 400 eV. Time step of 0.1 fs was employed. Mimicking the impacts of energetic electrons, initial kinetic energy was instantaneously assigned to the recoil atom. The atom was assumed to be displaced if it is moved by 4 {\AA} or more and the initial kinetic energy of the atom was taken as the threshold energy.

\section{Results} \label{sec:Results}

Fig.~\ref{Fig:Analysis} depicts ab-initio relaxed structures of bilayer (Fig.~\ref{Fig:Analysis}a) and monolayer (Fig.~\ref{Fig:Analysis}f) steps, respectively, together with a collection of simulated (Figs.~\ref{Fig:Analysis}a,b,f,g) and experimental (Figs.~\ref{Fig:Analysis}c,d,e,h,i) high-resolution (Figs.~\ref{Fig:Analysis} denoted by subscript 1) and medium-resolution (Figs.~\ref{Fig:Analysis} denoted by subscript 2) potential distributions at steps and edges comprising one (Figs.~\ref{Fig:Analysis}g-i) or two layer(s) (Figs.~\ref{Fig:Analysis}b-e). 

Most notably, the simulated structure of the bilayer step exhibits a strong deformation of the terminating layer structure in that the terminating bilayers are bend such to mutually interconnect at the step by forming an out-of-plane covalent bond (indicated by green isocharge surfaces). This strong deformation of the layers is not present in the simulated single step structure. Our ab-initio calculations, however, also show the formation of conducting states at the monolayer step similar to those obtained for the monolayer zig-zag edge \cite{Barone2008}). Subsequently, we conduct a detailed comparison of ab-initio simulation and experiment to verify (A) the structural and electronic reconstruction at h-BN steps, and (B) how the holographic potential data reflects this reconstruction. 

\begin{figure*}
    \centering
    \includegraphics[width=0.8\textwidth]{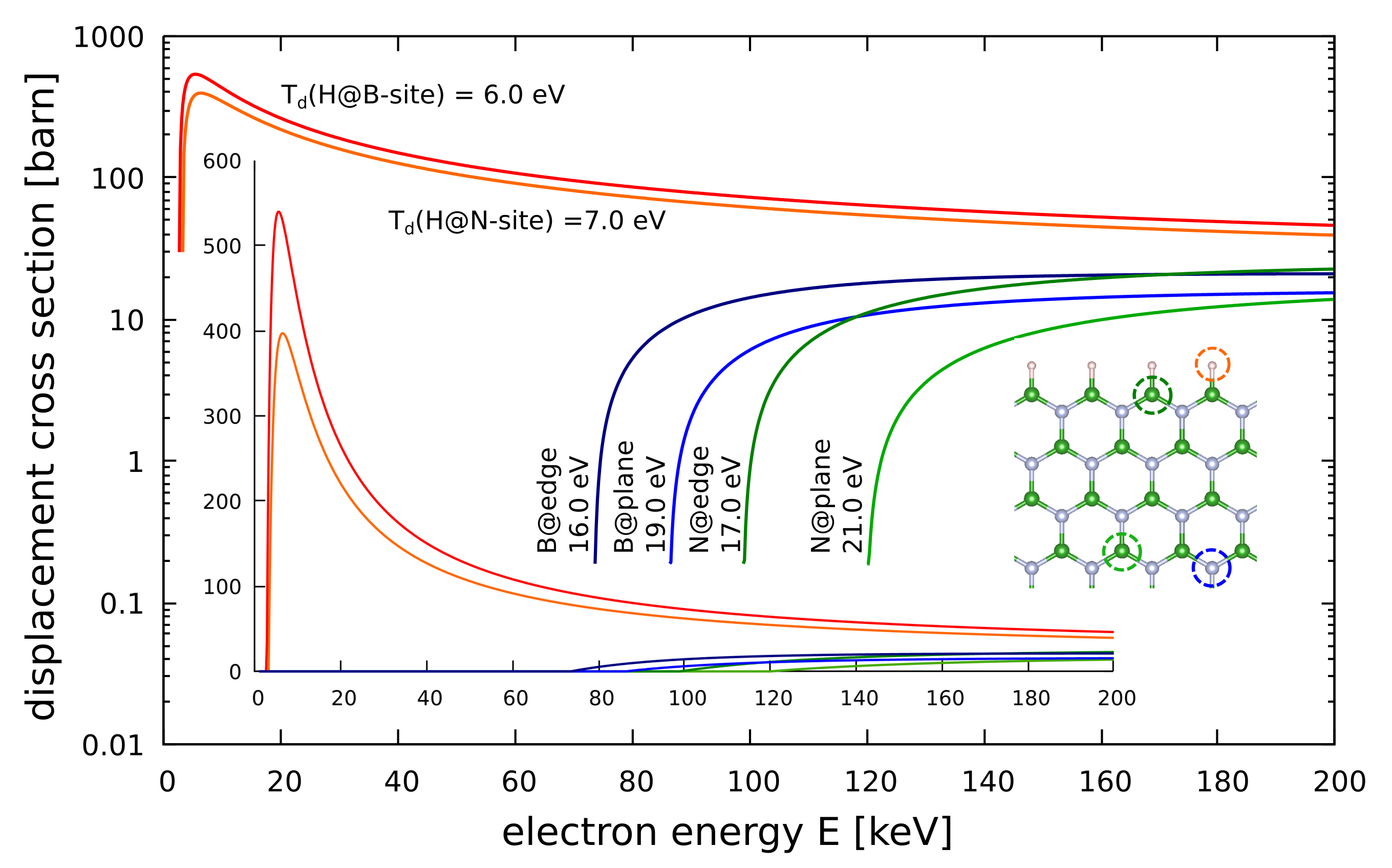}
    \caption{Electron energy dependent displacement cross sections for H/B/N atoms displaced from h-BN ribbons. For better comparability the ordinate scale on the main plot is logarithmic, the inset shows the linear scale. The inset image on the right illustrates the different target sites for the nitrogen terminated zigzag edge. Predominantly hydrogen atoms will be sputtered such that the edge will quickly become 'naked'.}
    \label{McF}
\end{figure*}

We first compare the high-resolution datasets, which provide a comprehensive picture of the in-plane structural changes (i.e., in-plane shift of the atomic positions). Projected potentials of bilayer steps in Fig. \ref{Fig:Analysis} with index 1 exhibit a subtle mutual lateral shift of the terminating atoms in adjacent h-BN layers (indicated by yellow arrows) both in the simulations and the experiment that are not present at monolayer steps. The presence of these shifts in the bilayer step may be explained by the loss of perfect stacking of adjacent layers due to the previously noted deformation of the bilayer step. Note, however, that while this observation indicates a profound difference between mono- and bilayer steps it does not allow to directly corroborate the predicted presence of the  folded bilayer step, i.e., the reconstruction both in in-plane and out-of-plane direction. 

In the DFT calculations, the latter is accompanied by the formation of a covalent bond between sp2 orbitals of terminating B and N atoms of the adjacent layers, as visible through the continuous electron density isosurface in Fig. \ref{Fig:Analysis}a. The corresponding delocalization of electrons lead to an increase of the simulated average projected potential of the bilayer step (Fig. \ref{Fig:Analysis}b$_2$), which is absent in the monolayer step (Fig. \ref{Fig:Analysis}g$_2$). Indeed, the holographically reconstructed medium resolution projected potentials agree very well with that theoretically predicted behaviour. The bilayer steps and edges exhibit a systematic increase of the potential at the edge (Fig. \ref{Fig:Analysis}c-e$_2$), which is not present at the monolayer steps and edges (Fig. \ref{Fig:Analysis}h-i$_2$, see also profiles in Fig.~\ref{Fig:Analysis}j). We can exclude that this effect originates from a systematic decoration/saturation by foreign atoms, in particular hydrogen, because of the previously noted in vacuo preparation of the samples. 

Moreover, our DFT-MD simulations indicate that even if the edges were decorated by H atoms, they
would have been quickly sputtered away by the electron beam. To prove that, we carried out
DFT-MD simulations, as described above, and obtained the following threshold displacement energies for H, B and N atoms: For B and N atoms in the pristine h-BN structure, the threshold energies were found to be  $T_d^B$ = 19 eV and $T_d^N$= 21 eV, which are close to the earlier calculated values \cite{Kotakoski2010}. Much smaller values of $~\sim 6/7$ eV were obtained for H atoms at B and N atoms. For the sake of completeness, displacement thresholds were also computed for B and  N
atoms at the edges.

The cross sections calculated using these displacement energies and employing the McKinley-Feshbach formalism are presented in Fig. ~\ref{McF}. It is evident that in the whole range of energies studied, the probability to displace H atom saturating dangling bonds at the edges of BN flakes is much higher than the probability to displace a B or N atom from the edges or the bulk of the system. Thus, even if the edges of h-BN flakes were saturated with H atoms, the edges would have been very quickly become ‘naked’.

Consequently, the information on electron delocalization encoded in the medium resolution projected potential data provides evidence for the strong out-of-plane structural and electronic reconstruction in the folded bilayer h-BN step. The monolayer step / edge, on the other hand, does not exhibit such additional bond formation and hence potential increase  (Fig.~\ref{Fig:Analysis}j).

\section{Summary} \label{sec:Summary}

We demonstrated that off-axis holographic determination of average projected potentials allows to directly image charge delocalization both perpendicular and along the projection (i.e., beam) direction. Holographically reconstructed medium- and high-resolution projected potential data in combination with ab-initio calculations of bilayer steps in h-BN revealed the formation of folded steps, i.e., a strong out-of-plane deformation of adjacent layers, which is absent in monolayer steps. As medium resolution holographic experiments can be readily performed in conventional TEMs down to low acceleration voltages, the technique offers new perspectives for the analysis of electronic properties of 2DMs. That particularly includes the study of edge and step reconstructions in various other 2DMs as well as defects of which, and of charge carrier accumulation at edges in polar 2DMs or 2D topological insulators (i.e., spin quantum hall systems).

\section{Acknowledgements}

The authors acknowledge funding from DFG SFB 1415, Project ID No. 417590517. We thank the Center for Information Services and High‐Performance Computing (ZIH) at TU Dresden for generous allocations of computer time. The authors gratefully acknowledge the Gauss Centre for Supercomputing e.V. (www.gauss-centre.eu) for funding this project by providing computing time through the John von Neumann Institute for Computing (NIC) on the GCS Supercomputer JUWELS\cite{juwels} at J\"ulich Supercomputing Centre (JSC). We furhter thank HLRS, Stuttgart, Germany, and TU Dresden Cluster “Taurus” for generous grants of CPU time.

\bibliographystyle{apsrev4-2}
\bibliography{2D}

\appendix*

\end{document}